\documentclass[english,showpacs,floatfix,12pt]{revtex4}
\usepackage[dvips]{graphicx}
\usepackage{longtable}
\usepackage{amsmath,amssymb}
\usepackage{dcolumn}
\usepackage{latexsym}
\usepackage{babel}
\usepackage[latin1]{inputenc}
\usepackage{color}
\usepackage[colorlinks]{hyperref}
\def\al{\alpha}
\def\vr{\varrho}
\def\kp{\kappa}

\def\pa{\partial}
\def\vf{\varphi}

\def\Om{\Omega}

\def\dl{\delta}
\def\Dl{\Delta}
\def\la{\lambda}
\def\La{\Lambda}
\def\th{\theta}
\def\sg{\sigma}

\def\f{{\cal F}}
\def\nn{\nonumber}
\def\diag{\mbox {diag}}

\def\wt{\widetilde}

\def\l{\left}
\def\r{\right}
\begin{document}
\title{\bf Functions of State for Spinor Gas in General Relativity}
\author{Ying-Qiu Gu}
\email{yqgu@fudan.edu.cn} \affiliation{School of Mathematical
Science, Fudan University, Shanghai 200433, China} 
\date{30th August 2017}

\begin{abstract}
The energy momentum tensor of perfect fluid is a simplified but
successful model in astrophysics. In this paper, assuming the
particles driven by gravity and moving along geodesics, we derived
the functions of state in detail. The results show that, these
functions have a little correction for the usual thermodynamics. The
new functions naturally satisfy the causal condition and consist
with relativity. For the self potentials of the particles we
introduce an extra function $W$, which acts like negative pressure
and can be used to describe dark matter. The results are helpful to
understand the relation and interaction  between space-time and
matter.

\end{abstract}
\maketitle

\section{Introduction}
\setcounter{equation}{0}

In astrophysics, the energy momentum tensor  is usually described by
perfect fluid model
\begin{eqnarray}
T^{\mu\nu}=(\rho+P)U^\mu U^\nu -P g^{\mu\nu}. \label{e1}
\end{eqnarray}
Theoretical analysis and experiments all show (\ref{e1}) is a
successful approximation. Some researchers such as Israel and
Stewart\cite{stew,Isra}, Carter\cite{Cart} and
Lichnerowicz\cite{Lich} disclose that, the energy momentum tensor
$T^{\mu\nu}$ includes abundant contents  for classical fluid theory,
which includes not only energy and momentum, but also heat flux,
spatial stress and viscosity\cite{and,his}. Furthermore, the
relativistic hydrodynamics with variational principle and Cartan's
exterior algebra was discussed in \cite{gour}.

The cooling mechanism of the expanding universe is one of the most
interesting problem of the students. The answers given by
pedagogical articles\cite{grv,tem3, tem4, tem5} are usually
$T\propto  a^{-1}$, which is derived from the classical
thermodynamics. In \cite{thm1,thm2,thm3,thm4,thm6}, the author
solved the thermodynamical relations according to the Gibbs' law
\begin{eqnarray}
\dl Q=d (\rho V) + P d V. \label{5}
\end{eqnarray}
The problem looks underdetermined due to (\ref{5}) including a
number of undetermined quantities. So we have to introduce some new
relations such as multi fluid with energy exchange, the apparent
horizon entropy, the decay of vacuum, to get a solution. As pointed
out in \cite{tem1}, these approaches did not convincing the students
and satisfying their curiosity. Dividing the velocity of particles
into Hubble velocity $v_h=r a'(t)$ and peculiar velocity $v_{pec}= a
r'(t)$, and then analyzing the geodesic of a particle, the author
qualitatively reached the conclusion $p\propto a^{-1}$. Then by
mass-energy relation $E^2=p^2+m^2$, he concluded that the cosmic
temperature should be $T\propto a^{-1}$ for the ultra-relativistic
gas but $T\propto a^{-2}$ for the non-relativistic gas.

As a matter of fact, free particles in gravitational field will
automatically move along geodesics, until they collide with each
other. So the functions of state of gases consistent with general
relativity should be naturally derived under background of gravity.
This is the purpose of the paper. We derive the functions in
microscopic point of view. From the calculation, indeed we can get
complete functions of state, which automatically satisfy the
requirement of relativity and remove the singularity from solutions
to the Einstein field equation\cite{gyq1}.

\section{Ideal gas in FRW space-time}
\setcounter{equation}{0}

For a dark spinor with self-interactive potential, in microscopic
view, the energy momentum tensor can be expressed by\cite{eng}
\begin{eqnarray}
T^{\mu\nu}=\sum_n(m_n u^\mu_n u^\nu_n +w_n g^{\mu\nu})\dl^3(\vec x
-\vec X_n)\sqrt{1-v^2_n},\label{e9}
\end{eqnarray}
where $m_n$ is the proper mass of the $n$-th spinor, $w_n>0$ is the
proper potential of self-interaction, $u^\mu_n$ 4-vector velocity,
$\vec v_n$ the usual 3-d speed, $\vec X_n(t)$ the central
coordinate. In the case of ideal gas, we have $w_n=0,(\forall n)$,
and then we have the energy-momentum tensor (\ref{e1})\cite{grv}. In
case $w_n>0$, the complete average formalism of energy-momentum
tensor should be
\begin{eqnarray}
T^{\mu\nu}=(\rho +P)U^\mu U^\nu +( W-P)g^{\mu\nu},\label{e28}
\end{eqnarray}
where $W$ is a new function of state reflect the self potential of
particles, which acts like negative pressure and is defined by
\begin{eqnarray}
W= \frac 1 V  \int_V \sum_n w_n \dl^3(\vec x -\vec
X_n)\sqrt{1-v^2_n} d V = \frac 1 V  \sum_{X_n\in V}  w_n
\sqrt{1-v^2_n}, \label{e290}
\end{eqnarray}

The functions of state in (\ref{e9}) and (\ref{e290}) including
relativistic factor $\sqrt{1-v^2_n}$, which can not calculate
directly. We solve the problem by the following method, which
naturally includes the interaction with gravity. To research the
thermodynamical properties of gas, we use piston and cylinder to
drive gas. In astrophysics, we have more ideal piston and cylinder,
that is the space-time with Friedmann-Robertson-Walker(FRW) metric,
which is absolutely adiabatic and reversible. The FRW metric drives
the gases homogeneously expanding and contracting as the scale
factor $a$ varies, and the results have general meanings according
to the principle of equivalence.

In the microscopic view, for the ideal gases and photons, the
particles are only driven by average gravity and move along
geodesics, and the collisions between particles can be treated as
instantaneous behavior. So all thermodynamic functions can be
rigorously solved according to dynamics and statistics. In this
section we set $w_n=0$ in (\ref{e9}) for simplicity.

For FRW space-time, we have the line element in conformal coordinate
system
\begin{eqnarray}
ds^2=a^2( t )\left(d t ^2-dr^2-{\cal S}^2(r)d\th^2-{\cal
S}(r)^2\sin^2\th d\vf^2\right), \label{1}
\end{eqnarray}
where
\begin{eqnarray} {\cal S}=\left \{ \begin{array}{ll}
  \sin r  & {\rm if} \quad  \kp =1,\\
   r   & {\rm if} \quad  \kp =0,\\
  \sinh r  & {\rm if} \quad  \kp =-1.
\end{array} \right. \label{2}
\end{eqnarray}
The energy conservation law $T^{\mu\nu}_{;\nu}$ in this case is
equivalent to (\ref{5}) while $\dl Q=0$, or manifestly
\begin{eqnarray} \frac {d(\rho  a^3)}{ d a}=-3 P
a^2.\label{4.1}\end{eqnarray} For a given equation $\rho=\rho(a)$,
we can solve the function $P=P(a)$ from (\ref{4.1}) or vice versa.

To solve geodesic in FRW space-time, we have the following result.

{\bf Lemma 1} {\em If the line element of the orthogonal subspace
has the following form,
\begin{eqnarray}
ds^2=\textbf{A}(t) d t^2 +\wt g_{\mu\nu}(t)d x^\mu dx^\nu,\label{6}
\end{eqnarray}
where $\textbf{A}$ and $\wt g_{\mu\nu}$ only depend on the
coordinate $t$, then the geodesic in this subspace can be solved by
\begin{eqnarray}
\frac {d x^\mu}{ds}=\wt g^{\mu\nu} C_\nu, \quad
\frac{dt}{ds}=\sqrt{\frac 1 {\textbf{A}}(1-\wt g^{\mu\nu} C_\mu
C_\nu)}, \label{7}
\end{eqnarray}
where  $C_\mu$ are constants, and $\wt g^{\mu\nu}\wt g_{\nu\al}
=\dl^\mu_\al$.}

Lemma 1 can be checked directly. For the FRW metric (\ref{1}), the
line element in the orthogonal subspace $(t,r)$ is given by
$ds^2=a(t)^2(dt^2- dr^2)$. According to Lemma 1, we have the
geodesic equation
\begin{eqnarray}
\frac {d}{ds} r =\frac C {a^2},\quad \frac {d}{ds} t =\frac 1
 {a^2} \sqrt{a^2+C^2}, \label{8} \end{eqnarray}
where $C$ is a constant only depends on the initial data. By
(\ref{8}) we get the drifting speed of a particle in usual sense
\begin{eqnarray}
v_n\equiv\frac{a dr}{a dt}=\frac {b_n}{\sqrt{a^2+b_n^2}},&~~~&
\sqrt{1-v_n^2}=\frac {a}{\sqrt{a^2+b_n^2}}.
\label{e19}\end{eqnarray} So the momentum of a particle $p= \frac
{m_nv}{\sqrt {1-v^2}}$ satisfies
\begin{eqnarray}
 p(t) a(t)=p(t_0)a(t_0),
\label{10}
\end{eqnarray}
where $m_n$ is the  proper mass of the particle. For the massless
photons, we can check that the wavelength $\la(t)$ satisfies $\frac
{\la(t)}{a(t)}\equiv \frac {\la_0}{a_0}$, so their momentum $p$ also
satisfy (\ref{10}). Although (\ref{10}) is derived in subspace-time
$(t,r)$, but it is suitable for all particles due to the symmetry of
the FRW metric.

The relation between momentum $p$ and the kinetic energy $K$ is
given by
\begin{eqnarray}p^2=K(K+2m).\label{11}\end{eqnarray}
By (\ref{10}) we have $ p^2_n=\frac {C_n}{a^2}$, where $C_n$ are
constants only depending on the initial data at $t=t_0$. Then on one
hand, for all particles we have the mean square momentum directly
\begin{eqnarray}
\bar p^2=\frac {C_0} {a^2}, \label{15}\end{eqnarray} where $C_0$ is
a constant only determined by initial data at $t_0$. One may argue
that (\ref{15}) is probably broken by the collision of the
particles. The following Lemma shows that (\ref{15}) holds in
statistical sense.

{\bf Lemma 2} {\em The mean square momentum of the ideal gas is
independent of the elastic collision of the particles.}

{\bf  Proof} For any elastic collision, we have momentum
conservation law $\vec p_1+\vec p_2=\vec P_1 +\vec P_2$, and then
\begin{eqnarray}
p_1^2+p_2^2=P_1^2+P_2^2+2(\vec P_1\cdot \vec P_2-\vec p_1\cdot \vec
p_2).\label{17}\end{eqnarray} Taking average for (\ref{17}), we have
\begin{eqnarray}
\bar p^2=\bar P^2 +\Dl.\label{18}\end{eqnarray} Since the elastic
collision is a reversible process, in statistical sense, we have the
exactly equal numbers of reversible process, so we also have
\begin{eqnarray}\bar P^2=\bar p^2 +\Dl.\label{19}\end{eqnarray}
Comparing (\ref{18}) with (\ref{19}), we have $\Dl=0$ and $\bar
p^2=\bar P^2$.  Since collision is finished instantaneously,
(\ref{15}) holds for all time $t$.

On the other hand, $\bar p^2$ can be calculated according to
statistical principle. Assuming the distribution of kinetic energy
$K$ of the particles is given by
\begin{eqnarray}
d {\cal P}=\f(K)dK, \label{e3}
\end{eqnarray}
then we have
\begin{eqnarray} \int _0^\infty d {\cal P}=1,~~~\int
_0^\infty K d {\cal P}=\frac 3 2 kT,~~~\int _0^\infty  K^2 d{\cal
P}=\frac 3 {2\sg} (kT)^2, \label{e4}
\end{eqnarray}
where $\sg$ is a constant reflecting the concrete distribution
function of particles.  In statistical mechanics, we usually use the
distribution functions of  momentum, which  is inconvenient for
calculation in the case of relativistic gases. Since the following
discussions have nothing to do with explicit function $\f(K)$, and
at most uses the second order moment, so the kinetic energy
distribution (\ref{e3}) is much convenient. In case of Maxwell
distribution, we have
\begin{eqnarray}
d {\cal P}&=& \exp\left(-\frac K{kT}\right) \sqrt{\frac {4K}{\pi
kT}}\frac {dK}{kT},\qquad \sg=\frac 2 5. \label{2.3}
\end{eqnarray}

By the moments (\ref{e4}) we have
\begin{eqnarray}
\bar p^2&=&\sum_n\int^\infty_0 \frac{1}{N}p^2_n \f(K_n)d K_n\nn\\
&=&\sum_n\int^\infty_0\frac{1}{N} K_n(K_n+2m_n) \f(K_n)d K_n\nn\\
&=&\sum_n kT\frac{1}{N}(\frac {3}{2\sg} kT+3m_n),\label{20}
\end{eqnarray}
where $N$ is the number of particles with mass $m_n$ in the volume
$V=\Om a^3$. Comparing (\ref{20}) with (\ref{15}), we get the
equation of $T(a)$ as follows
\begin{eqnarray}
kT=\frac{\sg\bar m b^2}{a(a+\sqrt{a^2+b^2})},\qquad a=\frac {\sg
\bar m b}{\sqrt{kT(kT+2\sg\bar m)}}, \label{e20}
\end{eqnarray}
where $ \bar m=\frac{ 1}{N}\sum_n m_n$ is the average mass of all
particles, and $b$ is a constant only depending on initial data.
Solving (\ref{e20}), we get

{\bf Theorem 3} {\em The temperature of ideal gases in FRW
space-time satisfies
\begin{eqnarray}
kT=\frac{\sg\bar m b^2}{a(a+\sqrt{a^2+b^2})}=\sg\bar
m\l(\sqrt{1+\frac{b^2}{a^2}}-1\r),\label{22}\end{eqnarray} where $b$
is constant determined by the initial data $a_0$ and $T_0$. $a(t)$
acts as intermediate parameter.}

By the theorem we find that, the cosmic temperature is different
from the results directly derived from classical thermodynamics. In
what follows we derive the relations between $\rho$ and $T$ as well
as the equation of state.

In microscopic point of view, the Lagrangian of FRW space-time
coupling with particles is given by\cite{grv,gyq5}
\begin{eqnarray}{\cal L}=\frac 1 {16 \pi G} (R-2\La)-\sum_n m_n
\sqrt{1-v_n^2}\dl^3(\vec x -\vec X_n),\label{25}\end{eqnarray} where
$\vec X_n$ is the coordinate of $n$-th particle, the scalar
curvature
\begin{eqnarray}R=6\frac {a''+\kp a}{a^3},\label{26}\end{eqnarray}
in which the prime stands for $\frac d {d t}$, and the drifting
speed of $n$-th particle in usual sense
$$\vec v_n=\left(v_r, v_\th,
v_\vf \right)_n=\left( \frac {adr}{ad t },\frac {a{\cal S} d\th}{ad
t }, \frac {a{\cal S}\sin\th d\vf}{a d t } \right)_n$$ are
independent variables related to $a$ for variation. Noticing that
$\dl^3(\vec x-\vec X_n)\propto a^{-3}$, by variation of $I=\int{\cal
L}a^4 dt d\Om$ with respect to $a$, we get
\begin{eqnarray}
a''+\kp a-\frac 2 3\La a^3=\frac{4\pi G}{3\Om}\sum_{X_n\in
\Om}m_n\sqrt{1-v^2_n},\label{27}
\end{eqnarray}
where $\Om$ is any given comoving volume with volume element $d\Om=
{\cal S}^2\sin \th dr d\th d\vf$, which is independent of $a$.
Substituting (\ref{e19}) into (\ref{27}), we get
\begin{eqnarray}
a''+\kp a-\frac 2 3\La a^3=\frac{4\pi G}{3\Om}\sum_{X_n\in \Om}\frac
{m_n a} {\sqrt{a^2+b_n^2}}.\label{28}
\end{eqnarray}
Multiply (\ref{28}) by $a'$ and integrate it, again by (\ref{e19})
we have
\begin{eqnarray}
a'^2+\kp a^2-\frac 1 3\La a^4&=&\frac{8\pi G}{3\Om}\sum_{X_n\in
\Om}\frac {m_n a} {\sqrt{1-v_n^2}}+C_1,\nn\\
&=&\frac{8\pi G}{3\Om}\sum_{X_n\in \Om}(K_n+m_n) a+C_1,\label{29}
\end{eqnarray}
where $C_1$ is a constant, for classical particles
$C_1=0$\cite{gyq5}.

By Friedmann equation and (\ref{22}),  making statistical average of
(\ref{29}) we get
\begin{eqnarray}
\frac{8\pi G}{3} \rho a^4&=&a'^2+\kp a^2-\frac 1 3\La a^4 \nn\\
&=&\frac{8\pi G}{3}\bar \rho \left( 1+\frac 3 2 \frac{\sg
b^2}{a(a+\sqrt{a^2+b^2})}\right)a^4,\label{30}
\end{eqnarray}
where $\rho$ is defined by (\ref{e1}) and $\bar \rho$ by the
following
\begin{eqnarray}\rho=\frac 1 V\sum_{X_n\in V}E_n=\frac 1 V\sum_{X_n\in V}(K_n+m_n),
\quad \bar\rho=\frac 1 V\sum_{X_n\in V}m_n=\frac \vr {a^3},
\label{31}\end{eqnarray} in which $\vr=\frac 1 \Om \sum_n m_n$ is
the comoving density independent of $a$. Comparing (\ref{30}) with
(\ref{e20}), we get

{\bf Theorem 4} {\em For the ideal gas in FRW space-time, the mass
density satisfies
\begin{eqnarray}\rho &=& \frac \vr{a^3} \left( 1+\frac
{3\sg} {2a} (\sqrt{a^2+b^2}-a)\right)=\bar\rho\left(1+\frac
3 2 \frac {kT}{\bar m}\right),\label{32}\\
\bar \rho & =& \vr_0 [kT(kT+2\sg\bar m)]^{\frac 3 2},
\label{32.1}\end{eqnarray} where $\vr_0$ is a constant.}

Substituting (\ref{32}) into (\ref{4.1}), we get

{\bf Theorem 5} {\em The equation of state for ideal gas in FRW
space-time is given by}
\begin{eqnarray}
P&=&\frac {\sg\vr b^2}{2a^4\sqrt{a^2+b^2}}=\frac {NkT} {V}
\left(1-\frac{ kT}{2(\sg\bar m+kT)}\right),\nn \\
&=& \frac {\vr_0 [kT(kT+2\sg\bar m)]^{\frac 5 2}} {2\bar
m(kT+\sg\bar m)}= \frac {\bar\rho kT(kT+2\sg\bar m)} {2\bar
m(kT+\sg\bar m)}. \label{33}\end{eqnarray}

The final expressions of function of state (\ref{32}) and (\ref{33})
are independent of metric $a$ which is only used as
`piston-cylinder' to drive the particles. So the state functions
(\ref{32}) and (\ref{33}) are generally valid for ideal gases in
general relativity.

Denoting $J=\frac {kT}{\bar mc^2}$, by (\ref{32}) and (\ref{33}) we
have asymptotic properties of equation of state(EOS) for the
particles
\begin{equation} P \dot = \left \{ \begin{array}{ll}
  P_0 {\rho}^\frac 5 3\l(1-\frac 1{2\sg}(5\sg+2)J\r), ~~& {\rm if~~} T\to 0 , \\
 \frac 1 3 \rho \l( 1+(\sg-\frac 2 3)J^{-1} \r), ~~& {\rm if~~} T\to
 \infty.
\end{array} \right. \label{eosa}
\end{equation}
The velocity of sound
\begin{eqnarray}
C_s\equiv c\sqrt{\frac {dP}{d\rho}}=\frac {\sqrt 3}
3\left(\frac{c^2J(2\sg+J)(5\sg^2+8\sg
J+4J^2)}{(\sg+J)^2[2\sg+(2+5\sg) J+4J^2]}\right)^{\frac 1 2}<\frac
{\sqrt 3} 3c.\label{2.4.1}
\end{eqnarray}
Obviously the EOS satisfies the increasing and causal conditions
which are consistent with relativity.

\section{Functions of state including self-potentials}
\setcounter{equation}{0}

In this section, we consider the case $w_n\ne 0$ in (\ref{e9}). By
(\ref{e19}), we have relation
\begin{eqnarray}
  \sqrt{1-v_n^2}  = \frac d {da}\frac
{a}{\sqrt{1-v^2_n}}=\frac 1 {m_n} \frac d {da} [ a( K_n+m_n )].
\label{e21}
\end{eqnarray}
This is the key relation to calculate the relativistic factor
$\sqrt{1-v_n^2}$. Substituting it into (\ref{e20}), and using
(\ref{22}) and (\ref{e4}), we get
\begin{eqnarray}
\sum_n \frac {m_n} N \int \sqrt{1-v_n^2} d{\cal P}& =& \frac d {da}
\l(\frac a N \sum_n \int ( K_n+m_n ) d{\cal P} \r) = \bar m-\frac
{3\sg
 \bar m kT}{2(\sg \bar m+kT)}, \label{e22}\\
\sum_n \frac{w_n} N \int \sqrt{1-v_n^2} d{\cal P}& =& \frac d {da}
\l(\frac a N \sum_n\frac {w_n}{m_n} \int ( K_n+m_n ) d{\cal P} \r) =
\bar w-\frac {3\sg
 \bar \mu\bar m  kT}{2(\sg \bar m+kT)}, \label{e23}
\end{eqnarray}
where the mean parameters are defined by
\begin{eqnarray}
\bar w= \frac 1 N \sum_n^N {w_n},\qquad \bar \mu= \frac 1 N \sum_n^N
\frac {w_n}{m_n}. \label{e24}
\end{eqnarray}
In the case $w_n>0$, by  (\ref{e23}) we have
\begin{eqnarray}
\sum_n w_n \int \sqrt{1-v_n^2} d{\cal P}>0, \qquad \sg \bar \mu <
\frac {2\bar w}{3\bar m}. \label{e25}
\end{eqnarray}
For the same kind particles  $\bar w= {\bar \mu}{\bar m}$, we find
$\sg<\frac 2 3$. If the scale of the space-time is stable and varies
slowly, e.g. in a galaxy, all the proper parameters such as $w_n$
and $m_n$ can be treated as constants.

If we define the mass-energy density $\rho$ and pressure $P$ of the
particles from (\ref{e9}) in micro form as usual\cite{grv}
\begin{eqnarray}
\rho \equiv \sum_n \frac {m_n}{\sqrt{1-v^2_n}} \dl^3(\vec x -\vec
X_n),~~~P \equiv \frac 1 3 \sum_n \frac {m_n v^2_n}{\sqrt{1-v^2_n}}
\dl^3(\vec x -\vec X_n),\label{e26}
\end{eqnarray}
then in mean sense we get
\begin{eqnarray}
\rho = \frac 1 V \int d{\cal P} \int_V \sum_n \left( K_n+m_n\right)
\dl^3(\vec x -\vec X_n)dV=\bar\rho \left( 1+\frac {3kT}{2\bar
m}\right). \label{e27}
\end{eqnarray}
Again we get (\ref{32}).  By (\ref{e22}) and (\ref{e26}), we have
\begin{eqnarray}
P = \frac 1 {3V} \int d{\cal P} \sum_{X_n\in V} \l( \frac {m_n}
{\sqrt{1-v_n^2}}+m_n\sqrt{1-v_n^2}\r)  =\bar\rho \frac {kT}{2\bar m}
\frac{(2\sg\bar m+ kT)}{(\sg\bar m+kT)}. \label{e30}
\end{eqnarray}
Again we get (\ref{33}). Similarly, by (\ref{e290}) and (\ref{e23})
we get $W$  in average form
\begin{eqnarray}
W = \frac 1 V  \int\sum_{X_n\in V}   w_n \sqrt{1-v^2_n}d{\cal P}
 =\bar\rho \left( \frac{\bar w}{\bar m}-\frac {3\bar \mu\sg
kT}{2(\sg \bar m +kT)}\right). \label{e29}\end{eqnarray}

The equations of state (\ref{e27}), (\ref{e30}) and (\ref{e29}) are
valid  for any ideal gases in local equilibrium. They have the
following dimensionless form,
\begin{eqnarray}
J &\equiv & \frac{kT}{\bar m c^2},\qquad \bar\rho = \vr_0 [J(J+2\sg)]^{\frac 3 2}, \label{ee30}\\
 \rho &=&\bar\rho \l(1+\frac 3 2 J\r)= \vr_0 [J(J+2\sg)]^{\frac 3
2}(1+\frac 3 2 J),
\label{e31}\\
P &=& \bar\rho \frac {J(J+2\sg)} {2(J+\sg)}=\vr_0 [J(J+2\sg)]^{\frac 5 2} \frac 1 {2(J+\sg)},  \label{e32}\\
W&=& \bar\rho \left( \mu -\frac {3\bar \mu\sg J}{2(J+\sg)}\right) =
\vr_0 [J(J+2\sg)]^{\frac 3 2} \left( \mu -\frac {3\bar \mu\sg
J}{2(J+\sg)}\right), \label{e33}
\end{eqnarray}
where $\vr_0$ is a constant depending on parameters $(\bar m,
b,\sg,\mu,\bar \mu)$, and $\mu=\frac {\bar w}{\bar m}$. Among the
functions of state only the temperature $kT$ or $J$ is independent
variable.

Equations (\ref{ee30})-(\ref{e33}) are based on the assumption that
the particles move along geodesic.  This is valid for $0\le w_n\ll
m_n$. In general cases,  these functions can be modified by the
following treatments. In (\ref{ee30})-(\ref{e33}), we take density
$\bar\rho$ or equivalently the volume $V$ as an independent state
function, and then derive the function $\bar\rho(J)$ according to
energy conservation law.

In the comoving coordinate system with the following Gaussian type
metric\cite{grv}
\begin{eqnarray}
g_{\mu\nu}=\diag\left( g_{00}, -\wt g_{ab}\right),\qquad (a, b)\in
\{1,2,3\}, \label{e35}
\end{eqnarray}
where $\wt g_{ab}$ is the spatial metric, for energy-momentum tensor
(\ref{e28}), we have

{\bf Theorem 6} {\em For the particles with energy-momentum tensor
(\ref{e28}), we have the following Gibbs-Duhem's law
\begin{eqnarray}
\dl Q =d [( \rho +W)V] + (P-W) d V, \label{e36}
\end{eqnarray} where $\dl Q$ denotes the heat received by the $N$ particles, and
\begin{eqnarray}
V=\sqrt{\wt g}\Dl x \Dl y \Dl z, \qquad ( \wt g= \det(\wt g_{a b}))
\label{e40}
\end{eqnarray}
is the micro spatial volume occupied by $N$ given particles.}

This can be checked as follows. We trace the motion of these
particles. In the adiabatic process, we have $\dl Q = 0$. Then by
$U_\mu T^{\mu\nu}_{~;\nu}=0$, we get the continuity equation for
 (\ref{e28}) as
\begin{eqnarray}
U^\mu\pa_\mu (\rho+W)+(\rho+P) U^\mu_{~;\mu}=0. \label{e37}
\end{eqnarray}
Denoting the proper time by $d\tau $, then $\frac d {d\tau
}=U^\mu\pa_\mu$,  (\ref{e37}) becomes
\begin{eqnarray}
0&=&\sqrt{|g|}\frac d{d\tau }(\rho +W)+(\rho
+P)\left(\sqrt{|g|}\pa_\mu U^\mu+
 \frac d {d\tau }\sqrt{|g|}\right),\nn \\
&=&\sqrt{g_{00}}\left(\frac d{d\tau }\left((\rho +W)\sqrt{\wt
g}\right)+(P-W)\frac d {d\tau }\sqrt{\wt
g}\right) \label{e38}\\
&&+(\rho +P)\sqrt{\wt g}\left(\sqrt{g_{00}}\pa_\mu U^\mu+ \frac d
{d\tau }\sqrt{g_{00}}\right),\nn
\end{eqnarray}
where $g= \det(g_{\mu\nu})=-g_{00}\wt g$. In the comoving system, we
have $U^\mu=(\sqrt{g^{00}},0,0,0)$ and $d\tau  = \sqrt{g_{00}} dt$,
then we get
\begin{eqnarray}
\sqrt{g_{00}}\pa_\mu U^\mu+ \frac d {d\tau
}\sqrt{g_{00}}=\sqrt{g_{00}}\pa_t \frac 1 {\sqrt{g_{00}}}+ \frac 1
{\sqrt{g_{00}}}\pa_t \sqrt{g_{00}}=0. \label{e39}
\end{eqnarray}
Multiplying  (\ref{e38}) by the comoving volume element $\Dl x \Dl y
\Dl z$ for the $N$ given particles, (\ref{e38}) gives (\ref{e36}) in
the case $\dl Q=0$. In the case of $\dl Q\ne 0$, (\ref{e36}) holds
due to energy conservation law.

Now we derive the relation $\bar\rho(J)$ from (\ref{e36}) for the
equilibrium process with $\dl Q=0$. For clearness, we use the the
static mass density $\bar\rho=\frac {N \bar m} {V}$ to replace $V$.
Substituting  (\ref{e27}), (\ref{e30}) and (\ref{e29}) into
 (\ref{e36}), we get dimensionless differential equation
\begin{equation}
\frac 1 {\bar\rho} \frac {d\bar\rho}{dJ}=\frac{3[(J+\sg)^2-\bar \mu
\sg^2]}{(J+\sg)[J^2+BJ-2\mu \sg]}. \label{e41}
\end{equation}
The solution is given by
\begin{eqnarray}
\bar\rho&=&\vr_0(J+\sg)^{\frac {3\bar\mu}{3\bar\mu+1}}\left( J^2+J
B-2\sg\mu \right)^{\frac 3 2\left(1-\frac
{\bar\mu}{3\bar\mu+1}\right)}\left( \frac{2J+B-A}
{2J+B+A}\right)^{\frac{\al}{2A}}, \label{e42}
\end{eqnarray}
where parameters are defined by
\begin{eqnarray}
B=(2+3\bar \mu)\sg-2\mu,\quad A=\sqrt{8\sg \mu+B^2},\quad \al=
3(1+4\bar\mu)[2\mu-3\bar\mu\sg]. \label{e43}
\end{eqnarray}
(\ref{e42}) shows how the internal potentials $w_n$ influence the
mass density. (\ref{e42}) reduces to (\ref{ee30}) if
$\mu=\bar\mu=0$.

\section{discussion and conclusion}
\setcounter{equation}{0}

\begin{enumerate}
\item
The above calculations  shows that functions of state consistent
with relativity should include the influences of gravity. The
energy-momentum tensor and geodesics  connect the macro concepts
with micro movements of particles.

\item  In the case of ideal gas with $\bar w=0$ at low
temperature, (\ref{e31})-(\ref{e32}) gives the equation of state for
the adiabatic monatomic gas
\begin{eqnarray}
P\dot = \rho J \dot = P_0 \rho^{\frac 5 3},\qquad (J\ll 1,~{\rm or}~
kT\ll \bar m), \label{e44}
\end{eqnarray}
which is identical to the empirical law in thermodynamics. When
$J\gg 1$, we have $P\to \frac 1 3 \rho$, so the adiabatic index is
not a constant for large range of temperature due to the
relativistic effect. These results show the validity of
(\ref{e31})-(\ref{e33}) and the consistence with normal
thermodynamics. \item By (\ref{e31}), letting $J\to \infty$ or $\bar
m\to 0$, we get the Stefan-Boltzmann's law $\rho \propto T^4$. This
means that the above results automatically include photons, and the
Stefan-Boltzmann's law is also valid for the ultra-relativistic
particles.

\item In general relativity, all processes occur
automatically, and $\vr_0$ is independent of any practical process.
Of course, $\vr_0$ is related to the property of particles.
Furthermore, equation of state (\ref{e31})-(\ref{e32}) provides a
singularity-free stellar structure in thermal
equilibrium\cite{gyq1}.

\item In the case $\bar w > 0$, the motion of the particles will slightly deviate from the
geodesic. By (\ref{e42}) we find, $\bar\rho=0$ leads to $J = \frac
{2\sg \mu}{B}\approx \mu $, which means the zero temperature can not
reach. The physical reason for such conclusion is unclear.
\end{enumerate}

\section*{Acknowledgments}
The author is grateful to Prof. Ta-Tsien,  Prof. Tie-Hu Qin and
Prof. Ji-Zong Li for  encouragement.

\end{document}